**Beam experiments for reactive ion etching of silicon (Si)-based materials by silicon halide ions**

Kazuhiro Karahashi[1], Tomoko Ito[1], and Satoshi Hamaguchi[1]

[1]*Center for Atomic and Molecular Technologies, Osaka University, Yamadaoka 2-1, Suita 565-0871, Japan*
[2]*TEL Technology Center Tsukuba, Tokyo Electron Ltd.*

**Abstract**

Etching yields of Si, $SiO_2$, and $Si_3N_4$ have been determined for silicon ion ($Si^+$), halogen ions ($F^+$, $Cl^+$, and $Br^+$) and silicon halide ions ($SiF^+$, $SiF_3^+$, $SiCl^+$, $SiCl_3^+$, $SiBr^+$, and $SiBr_3^+$) irradiation in 300 to 1000 eV using a mass-selected ion beam apparatus that can irradiate a single species ion to sample surfaces under an ultra-high vacuum condition. Si+ irradiation below 1000eV deposits silicon atoms on Si, $SiO_2$, and $Si_3N_4$ surfaces. The etching yields of silicon tri-halide ions ($SiF_3^+$, $SiCl_3^+$, and $SiBr_3^+$) above 1000 eV are larger than those of halogen ions, respectively, and these etching yields depend on the incident ion energy and the etching material (especially $Si_3N_4$). At low incident energy, silicon mono-halide ions ($SiF^+$, $SiCl^+$, and $SiBr^+$) deposit silicon on substrates, and the etching threshold energy depends on the halogen species. This information contributes to a database of experimental values needed to increase the precision of an etching process and a profile simulator.

# I. INTRODUCTION

Reactive ion etching (RIE) by halogen-based plasmas is widely used for Si, $SiO_2$, and $Si_3N_4$ etching in semiconductor manufacturing processes [1]-[5]. As semiconductor devices continue to miniaturize, a better understanding of basic reactions of etching and/or deposition processes on substrate surfaces has become more important than before for finer controls of device structures in the manufacturing processes [6][7]. It is known that chlorine and HBr plasmas exhibit high etching rates, high selectivity, and high etching anisotropy on Si surfaces, while fluorocarbon plasmas containing fluorine are widely used for etching oxide and nitride films. Silicon halides, which are etching by-products released from the surface during these etching processes, influence the etching reaction by reinjecting onto the surface. In recent 3D NAND development, the formation of high-aspect-ratio contact holes has become a critical technology. It is necessary to etch alternating layers of $SiO_2$ and $Si_3N_4$, stacked hundreds of layers high, with an aspect ratio of 100 or more. Therefore, high-aspect-ratio etching technology is essential. This technology is also key for the transition to 3D DRAM, which is currently attracting significant attention. In high-aspect-ratio patterns, Knudsen diffusion reduces the number of radicals reaching the etching front at the bottom of the pattern to approximately 1%. In conventional RIE processes, the etching reaction depends strongly on the ratio of radicals to ion flux; however, in high aspect ratio etching, reactive ions play a dominant role in the etching reaction [8][9]. Therefore, to develop highly controllable Si, $SiO_2$, and $Si_3N_4$ etching processes using halogen-based plasma, it is important to clarify the etching characteristics of silicon halide ions. Beam experiments are a very useful tool for understanding interactions of individual species contained in a plasma with the surface [10]-[17]. We reported on the characteristics of silicon etching by silicon chloride ions using a mass-selected ion beam system. We found that the etching reaction of silicon-chloride ions depends on the ion species. At low energy, deposition of silicon atoms contained in the $SiCl^+$ ion occurs [17]. In this study, we examine etching and/or deposition reactions of Si, $SiO_2$, and $Si_3N_4$ substrates using a mass-selected ion beam system. This approach can clarify the roles of halogen ions ($F^+$, $Cl^+$, and $Br^+$), silicon mono-halide ions ($SiF^+$, $SiCl^+$, and $SiBr^+$), and silicon tri-halide ions ($SiF_3^+$, $SiCl_3^+$, and $SiBr_3^+$).

## II. EXPERIMENT

A schematic diagram of a mass-selected ion beam system used in this study is given in Fig. 1 [15][16]. It consists of an ion-beam-source chamber with a Freeman-type ion source and a scattering chamber. Since there are three differentially pumping stages between the ion source and scattering chamber, the scattering chamber was maintained at ultrahigh vacuum (UHV) conditions below $5x10^{-7}$ Pa. Therefore, all experiments were conducted under conditions where the results were not affected by neutral radicals from ion sources. Various ions were generated in arc plasma using Ar-diluted $SiF_4$, $SiCl_4$, or $SiBr_4$ gas and were extracted at 25 keV from the ion source. Si+ ion, halogen ions ($F^+$, $Cl^+$, and $Br^+$), silicon mono-halide ions($SiF^+$, $SiCl^+$, and $SiBr^+$), and silicon tri-halide ions($SiF_3^+$, $SiCl_3^+$,and $SiBr_3^+$) for sample irradiation were selected with a 90° mass-analyzing electromagnet, passed through beam optics, and decelerated to a specified incident ion energy between 300 and 1000 eV just before they reached the sample. All samples were set to the normal incident of the ion beam in the scattering chamber. The ion dosage irradiating the samples was calculated from the ion current measured with a Faraday cup with a diameter of 1mm placed at the sample-irradiation position. Typical ion current density was 6-20uA/cm2, and typical beam size was 2-4 mm full width at half maximum.

Samples were prepared from Si(100) wafers. poly-Si or $Si_3N_4$ films were deposited by chemical vapor deposition (CVD), and $SiO_2$ films were thermally grown to a thickness of approximately 200 nm. After ion-beam irradiation through a stencil line-and-space mask (0.1 mm linewidth), etch depths were measured using a surface profiler. Etching yields were defined as the ratio of the number of Si atoms removed from the surface to the number of incident ions. The removed Si atoms were estimated from the measured etch depth and the densities of poly-Si (1.9 g/cm$^3$), $SiO_2$ (2.2 g/cm$^3$), and $Si_3N_4$ (2.1 g/cm$^3$). The ion dose was calculated from the irradiation time and the ion current. When irradiation resulted in deposition rather than etching, the yield per incident ion was calculated assuming deposition of a silicon film.

## III. RESULTS AND DISCUSSION

### A. Etch yield of Si

Figure 2 shows the ion energy dependence of the Si etching yield for (a) F+, SiF+, and $SiF_3^+$; (b) $Cl^+$, $SiCl^+$, and $SiCl_3^+$; and (c) $Br^+$, $SiBr^+$, and $SiBr_3^+$. The white circles represent yields from halide ions, the gray circles represent yields from mono-halides, and the black circles represent yields from tri-halides. The etching yield for each ion increases with increasing incident ion energy, indicating that ion energy promotes the etching reaction; however, the dependence on ion energy varies significantly depending on the ion species. For halide ion implantation, which involves single-atom ions, steady-state etching occurs in the 300–1000 eV range. At 1000 eV, the etching yields for fluorine, chlorine, and bromine are nearly identical; however, when the incident energy is less than 500 eV, the etching yield of the Br+ ion is lower than that of the $F^+$ and $Cl^+$ ions. This trend at energies below 500 eV is attributed to differences in the chemical reactivity of each halogen species. Specifically, the bond energy between fluorine and the silicon atoms—which serve as substrate atoms in silicon fluoride, silicon chloride, and silicon bromide compounds—is lowest for fluorine and increases in the order of chlorine and then bromine. Consequently, fluorine atoms spontaneously etch the silicon surface at room temperature [18]–[25], silicon chloride desorbs at a lower temperature than silicon bromide [26]. The etching yields at 1000 eV for the tri-halide ions ($SiF_3^+$, SiCl3+, and $SiBr_3^+$) are higher than those for the halide ions ($F^+$, $Cl^+$, and $Br^+$). In the case of mono-halide ions ($SiF^+$, $SiCl^+$, and $SiBr^+$), the etching yield is lower than that of halide ions, and at low incident energies, silicon is deposited on the substrate. The threshold energy of $SiBr^+$ is approximately 500 eV, which is higher than that of $SiF^+$ and $SiCl^+$. Figure 3 shows the change in etching yield when $Si^+$ ions are irradiated onto silicon. Even at 1000 eV, etching does not occur, and silicon atoms are deposited on the surface.

Generally, when polyatomic molecular ions with an incident energy of 100 eV or more are irradiated onto a surface, it is believed that the molecules dissociate into individual atoms before penetrating into the solid. When tri-halide ions collide with the surface, the kinetic energy of the incident ions is distributed according to atomic mass, causing the ions to dissociate into a silicon atom and three halogen atoms. Figure 4 compares the yield of tri-halide ions with the combined yield of silicon atoms and three halogen atoms. The yields of silicon atoms and halogen ions at various incident energies resulting from dissociation were calculated by interpolating experimental data for the individual yields. The experimental yields for tri-halide ions are larger than the values estimated from the yields of individual atomic ions. The magnitude of this increase is greatest for fluorine, followed by chlorine and bromine. In particular, there is a significant difference between the etching yield of the $SiBr_3^+$ ion and the estimated etching yield; the etching yield of $SiBr_3^+$ is more than three times the estimated value. Since the collision cascades induced by tri-halide ions decomposing into individual atoms and penetrating the solid occur within a narrow region, the collision cascades for each atom are thought to be strongly correlated. As a result, compared to the sum of the etching yields when the collision cascades of silicon atoms and halogen atoms are independent, each collision cascade occurs within a narrow region. These results suggest that the reaction of polyatomic ions at low incident energies is a complex process involving chemical reactions.

## *B.* Etch yield of $SiO_2$ and $S_3N_4$

Figure 5 shows how the etching yield of silicon oxide depends on ion energy for (a) $F^+$, $SiF^+$, and $SiF_3^+$; (b) $Cl^+$, $SiCl^+$, and $SiCl_3^+$; and (c) $Br^+$, $SiBr^+$, and $SiBr_3^+$. The etching yield of silicon oxide for each ion species is lower than that for silicon or silicon nitride. At 1000 eV, the etching rate of silicon oxide by halogen atomic ions is one-third that of silicon. This rate of decrease in etching rate is comparable to that observed with physical sputtering rates, such as those using $Ar^+$ ions [12]; however, at energies below 500 eV, the rate of decrease in etching rate becomes more pronounced. This is because atomic fluorine and chlorine do not easily etch silicon oxide substrates like silicon does [27]. Figure 6 shows how the etching rate of silicon nitride depends on ion energy. At 1000 eV, the etching rate of silicon nitride lies between that ofsilicon and silicon oxide. However, the etching yield by low-energy tri-halide ions (500 eV or less) decreases abruptly as the ion energy decreases, and for mono-halide ions, a silicon atom deposition reaction proceeds.

Figure 7 plots the thickness changes of silicon, silicon oxide, and silicon nitride as a function of ion dose following $Si^+$ irradiation. Silicon was deposited on all substrates, and the deposition rate increased as the incident ion energy decreased. The thickness change on silicon increased linearly with increasing ion dose, whereas the thickness changes on the silicon nitride and silicon oxide surfaces at an ion dose of $4 \times 10^{17}$ ions/cm$^2$ were greater than those on the silicon surface. At deposition layer thicknesses exceeding the ion penetration depth, the deposition rate is not affected by the substrate material; therefore, in the initial stage of ion irradiation, silicon atoms deposit rapidly on silicon nitride and silicon oxide surfaces compared to the silicon surface. Consequently, the silicon atoms contained in the silicon halide ions inhibit the etching reaction. In the low-incidence-energy region, the etching yield of tri-halide ions is suppressed by the silicon atoms contained in the ions, so the etching yield decreases rapidly as the ion energy decreases. When $CF_3^+$ ions are irradiated onto a silicon nitride film, carbon tends to deposit at low energies of 500 eV or less, similar to $SiF_3^+$ ions [28]. Furthermore, the etching yield of $CF_3^+$ ions increases sharply at energies of 1000 eV or higher, similar to that of $SiF_3^+$ ions. This suggests that carbon deposits more readily on silicon nitride films than on silicon films. It is thought that the reason such silicon oxide films have different properties from silicon and silicon nitride films is due to the chemical effects of oxygen and nitrogen; however, the reaction mechanism has not been fully elucidated based on these experimental results alone. In the future, it will be necessary to proceed with measurements of reaction products and other factors.

## IV. CONCLUSIONS

In this study, we investigated the etching characteristics of Si, $SiO_2$, and $Si_3N_4$ using a mass-selective ion beam system with halide ions ($F^+$, $Cl^+$, $Br^+$), mono-halide ions ($SiF^+$, $SiCl^+$, $SiBr^+$), and tri-halide ions ($SiF_3^+$, $SiCl_3^+$, $SiBr_3^+$). The results revealed that the etching yield per halogen atom for tri-halide ions is significantly higher than that for mono-atomic ions, indicating that the etching reaction proceeds more efficiently. This effect becomes more pronounced with increasing incident energy and is greater for bromine than for fluorine or chlorine. Furthermore, in the low-energy region (500 eV or less), deposition reactions of silicon films are induced; however, the initial process depends heavily on the substrate material, and it was found that silicon atoms tend to remain more readily in oxide and nitride films than in silicon substrates. These quantitative data are expected to contribute to the elucidation of etching processes and the improvement of the accuracy of profile simulators.

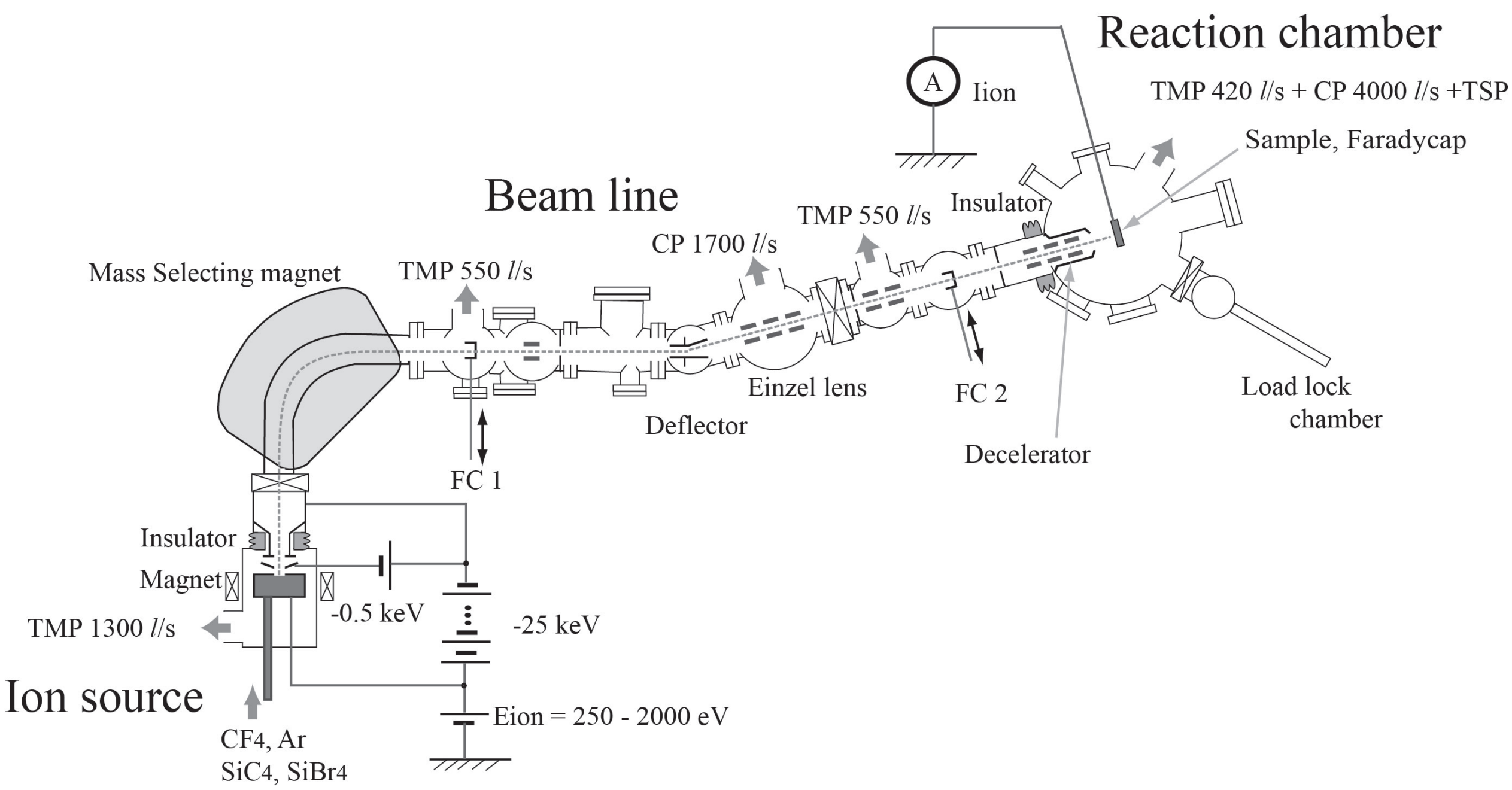

FIg.1 A schematic bird's-eye view of the mass-selected ion beam system used in this study.

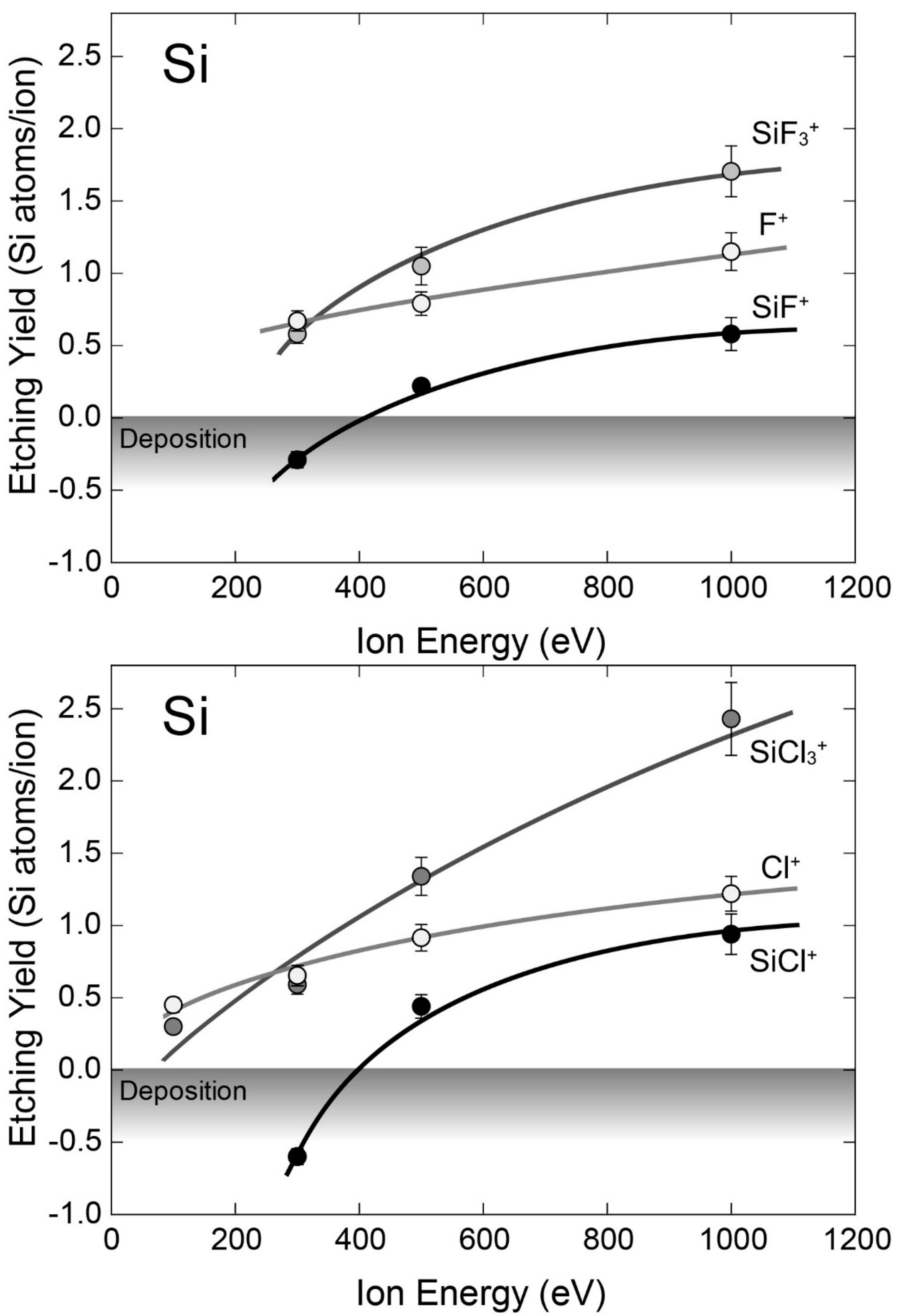
Si
SiF3+
F+
SiF+
Deposition
Etching Yield (Si atoms/ion)
Ion Energy (eV)
Si
SiCl3+
Cl+
SiCl+
Deposition
Etching Yield (Si atoms/ion)
Ion Energy (eV)

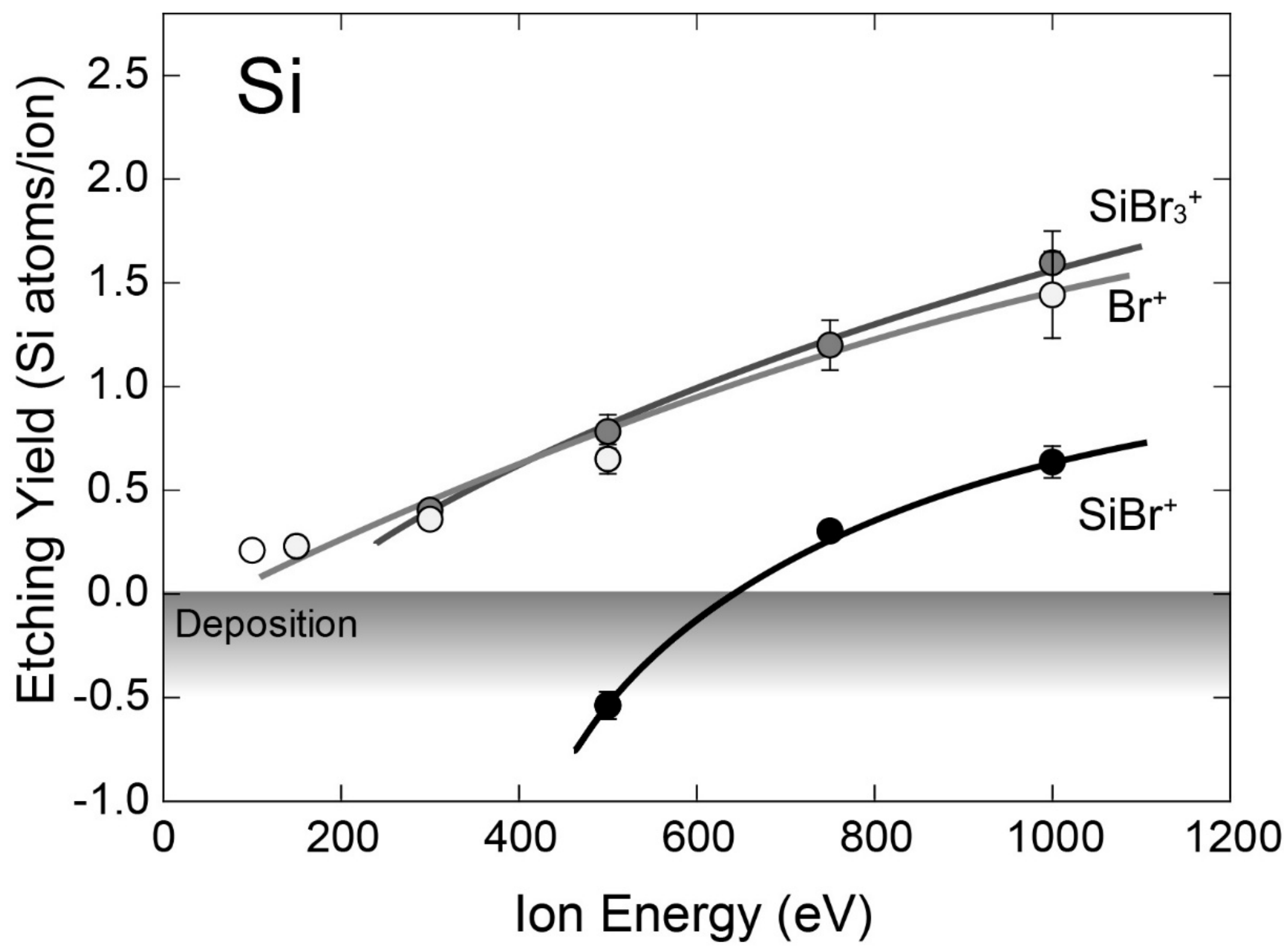

Fig.2 (a) Etching yields of silicon by $F^+$, $SiF^+$, and $SiF_3^+$ ions as functions of the ion incident energy. In this energy range, incident $Si^+$ ions cause deposition. The angle of incidence is normal to the sample surface in this study. (b) Etching yields of silicon by $Cl^+$, $SiCl^+$, and $SiCl_3^+$. (c) Etching yields of silicon by $Br^+$, $SiBr^+$, and $SiBr_3^+$ ions. The solid curve is a guide to the eye.

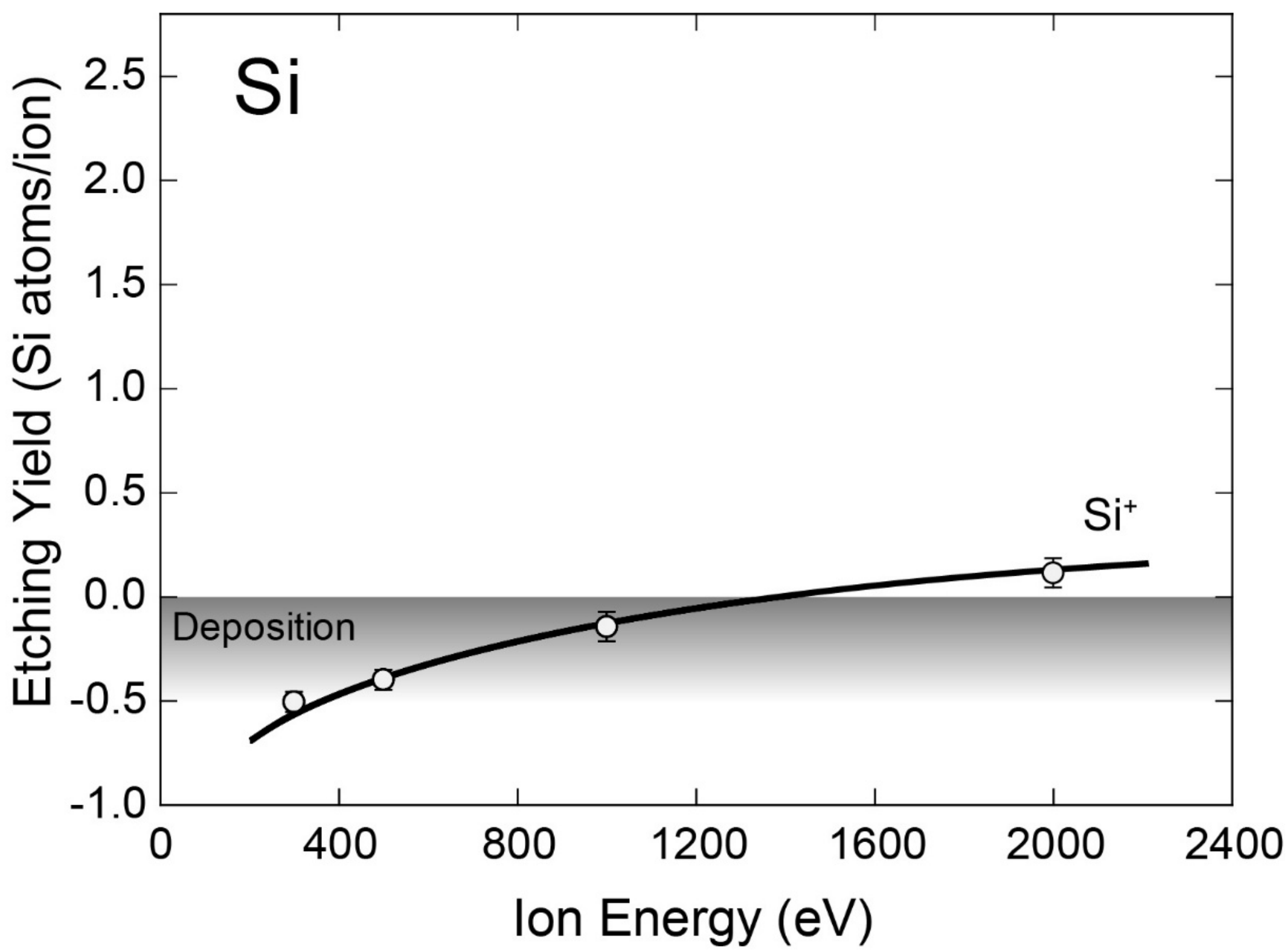

Fig.3 Etching yields of silicon by $Si^+$ ions as functions of the ion incident energy. Below 1000 eV, incident $Si^+$ ions cause deposition.

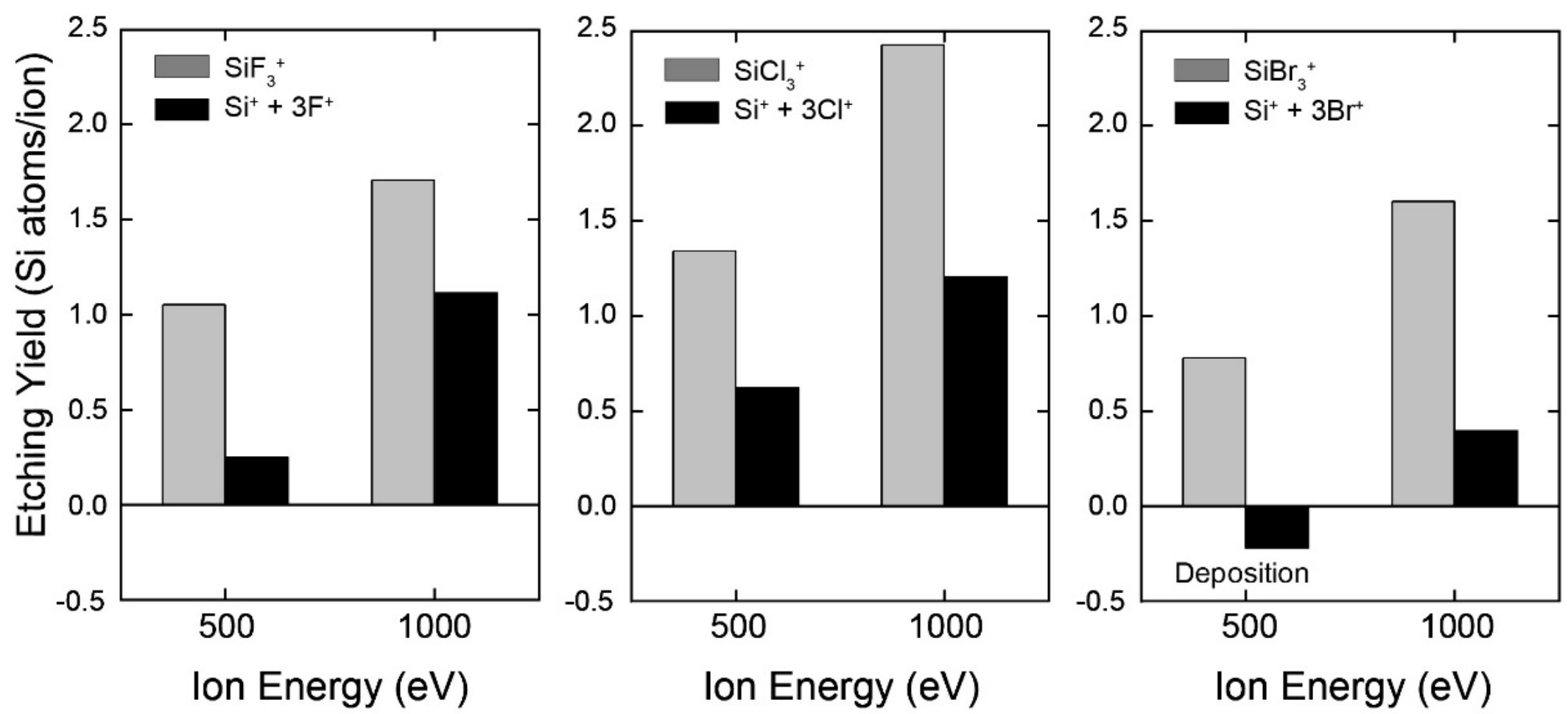

Fig.4 Etching yields of silicon by tri-halide ion; (a)$SiF_3^+$, (b)$SiCl_3^+$, (c)$SiBr_3^+$, and estimated yields from the yields of silicon and three halogen ions.

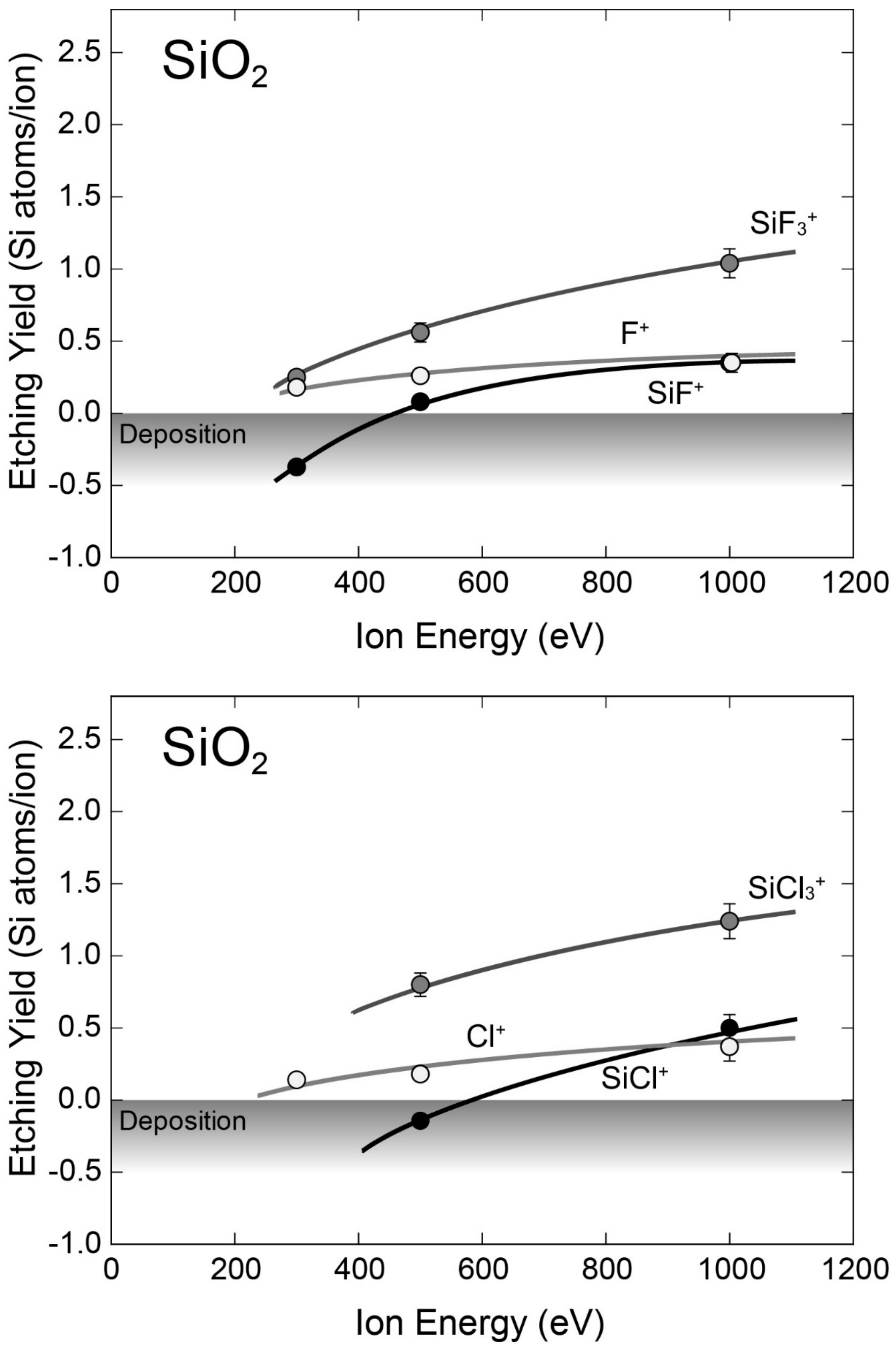
SiO2
SiF3+
F+
SiF+
Deposition
Etching Yield (Si atoms/ion)
Ion Energy (eV)
SiO2
SiCl3+
Cl+
SiCl+
Deposition
Etching Yield (Si atoms/ion)
Ion Energy (eV)

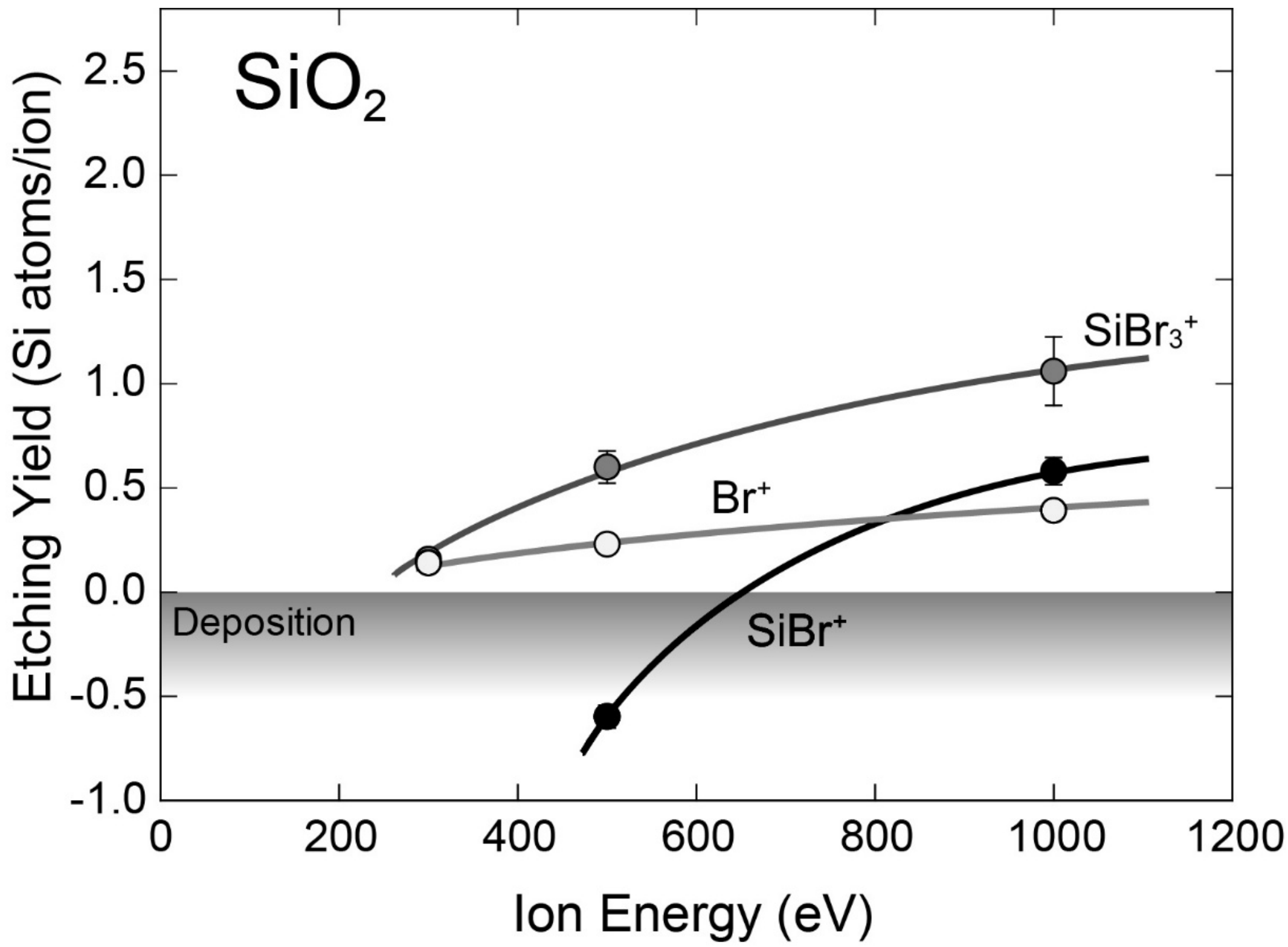

Fig.5 (a) Etching yields of silicon-oxide by $F^+$, $SiF^+$, and $SiF_3^+$ ions as functions of the ion incident energy. The angle of incidence is normal to the sample surface in this study. (b) Etching yields by $Cl^+$, $SiCl^+$, and $SiCl_3^+$ . (c) Etching yields by $Br^+$, $SiBr^+$, and $SiBr_3^+$ ions. The solid curve is a guide to the eye.

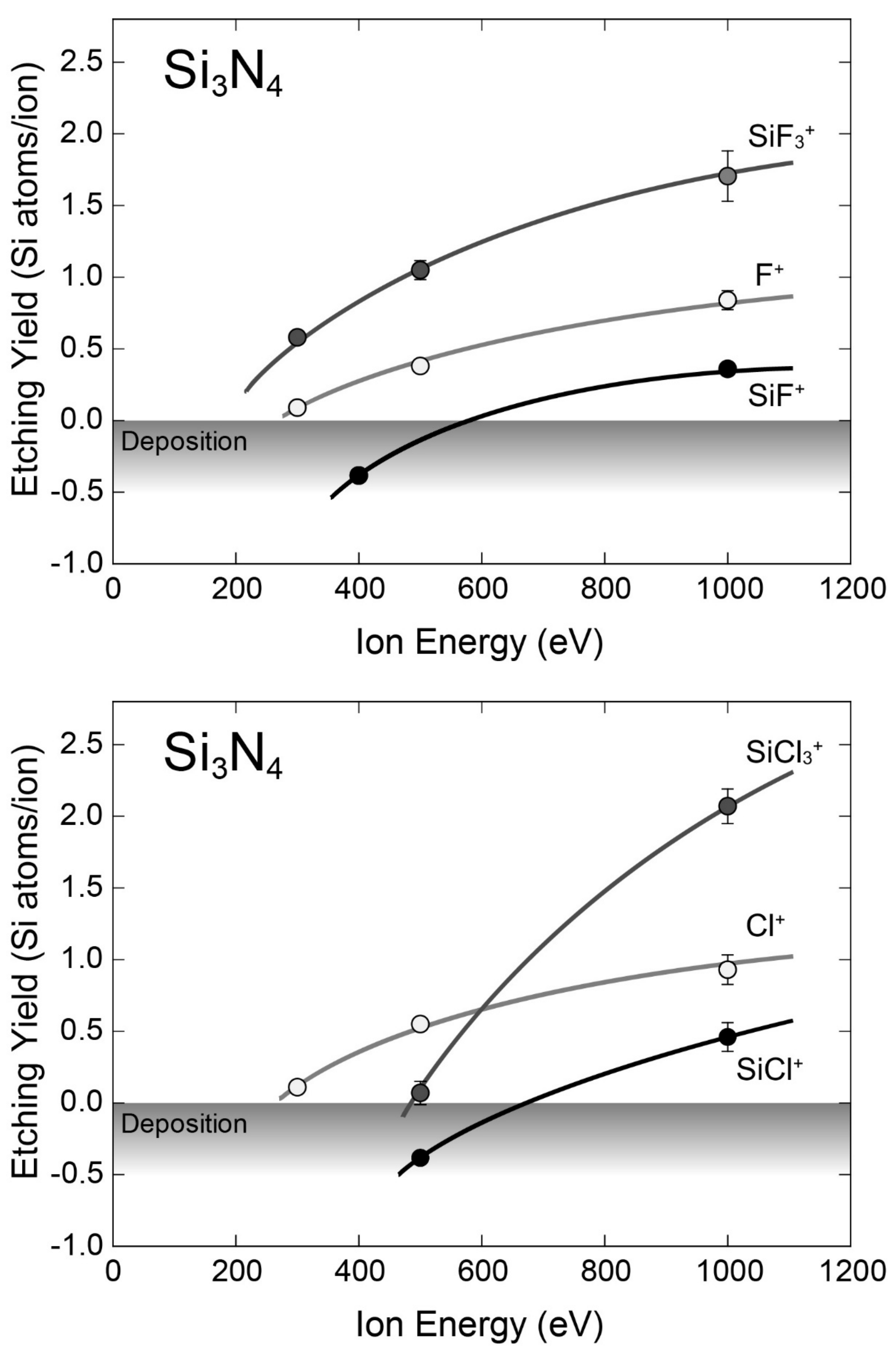

$Si_3N_4$
$SiF_3^+$
$F^+$
$SiF^+$
Deposition
Etching Yield (Si atoms/ion)
Ion Energy (eV)
2.5
2.0
1.5
1.0
0.5
0.0
-0.5
-1.0
0
200
400
600
800
1000
1200
$Si_3N_4$
$SiCl_3^+$
$Cl^+$
$SiCl^+$
Deposition
Etching Yield (Si atoms/ion)
Ion Energy (eV)

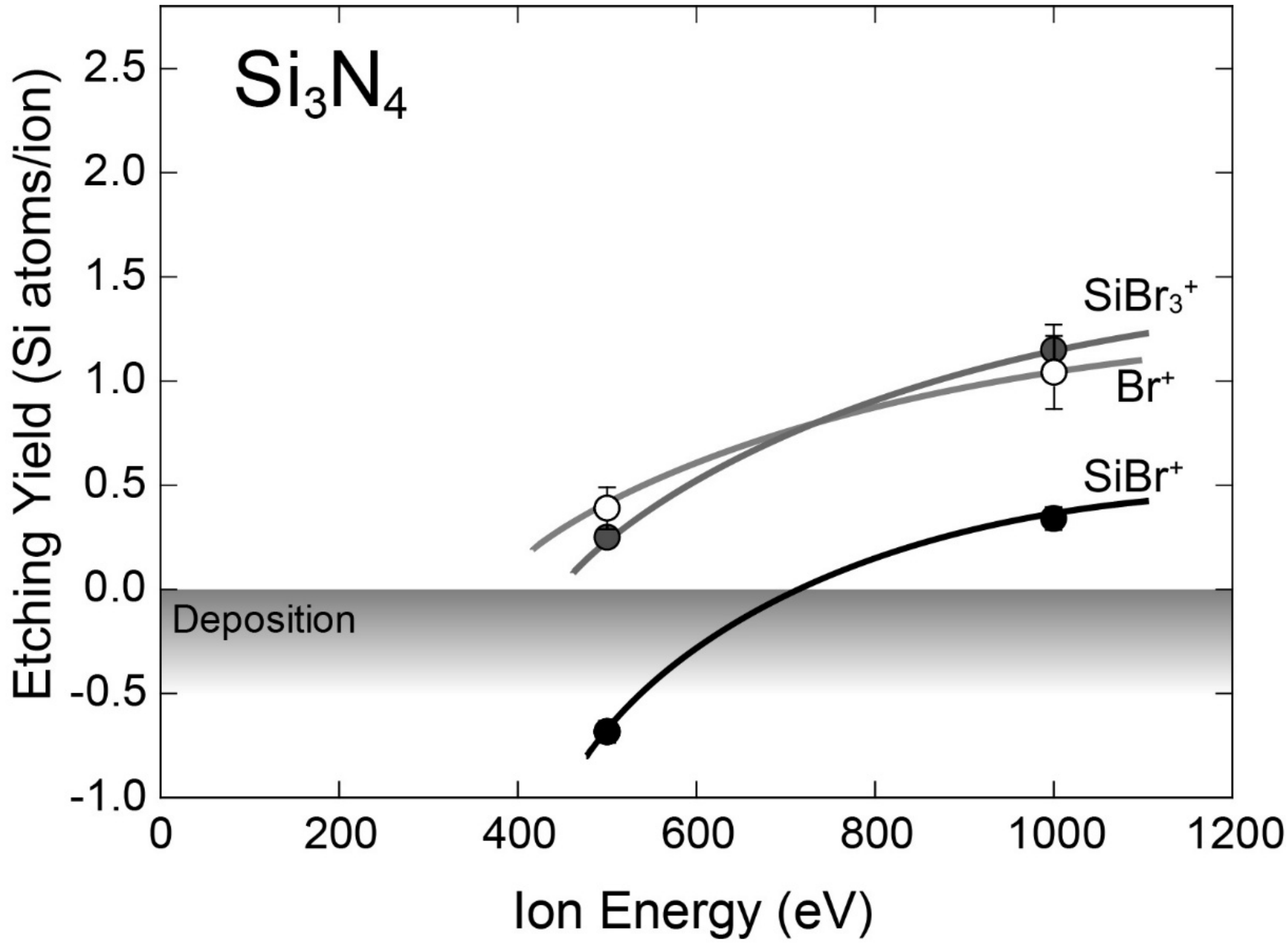

Fig.6 (a) Etching yields of silicon-nitride by $F^+$, $SiF^+$, and $SiF_3^+$ ions as functions of the ion incident energy. The angle of incidence is normal to the sample surface in this study. (b) Etching yields by $Cl^+$, $SiCl^+$, and $SiCl_3^+$. (c) Etching yields by $Br^+$, $SiBr^+$, and $SiBr_3^+$ ions. The solid curve is a guide to the eye.

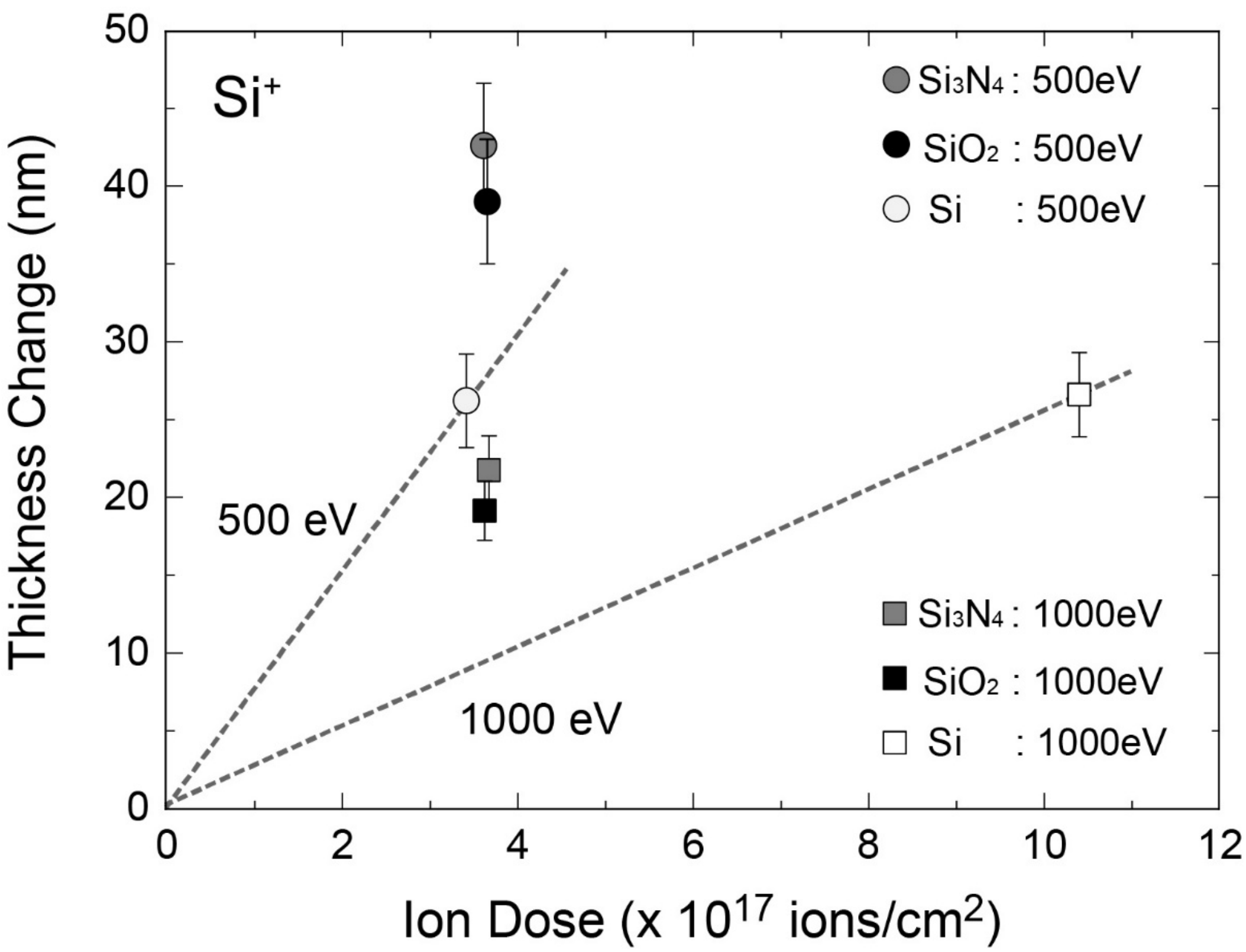

Fig.7 The thickness of a silicon film deposited by 500 and 1,000 eV $Si^+$ ion irradiation as a function of the ion dose on silicon, silicon-oxide, and silicon-nitride films.